\documentstyle[version2,aps,preprint]{revtex}
\begin{document}
\draft
\begin{title}
Interactions for odd-$\omega$ gap singlet superconductors
\end{title}

\author{Elihu Abrahams}
\begin{instit}
Serin Physics Laboratory, Rutgers University, P.O. Box
849, Piscataway, New Jersey 08855
\end{instit}
\author {Alexander Balatsky}
\begin{instit}
Theory Division, T-11, Los Alamos National Laboratory, Los
Alamos, New Mexico 87545\\
and Landau Institute for Theoretical Physics, Moscow,
Russia.
\end{instit}
\author {J.R. Schrieffer}
\begin{instit}
Department of Physics, Florida State University, Tallahassee, Florida 32310
\end{instit}
\author {Philip B. Allen}
\begin{instit}
Department of Physics, State University of New York, Stony Brook, New York
11794-3800
\end{instit}
\begin{abstract}
A  class of singlet superconductors with a gap function  $\Delta({\bf k},
\omega_n)$ which is {\it odd} in both momentum and   Matsubara frequency
was proposed recently \cite{ba}.  To show an instability in the {\it odd}
gap channel, a model phonon  propagator was used with the $p$-wave
interaction strength larger than the $s$-wave. We argue that the
positive   scattering matrix element entering the Eliashberg equations
leads to a  constraint on the relative strength of $p$- and $s$-wave
interactions which inhibits odd pairing. However, a general spin
dependent electron-electron interaction can satisfy all constraints and
produce  the odd singlet gap. A possibility which may lead  to an odd
gap is a strongly antiferromagnetically correlated system,  such as a
high-$T_c$ material.
\end{abstract}
\narrowtext

Balatsky and Abrahams \cite{ba} recently proposed a class of
singlet superconductors which exhibit unconventional symmetry of  the
pairing order parameter $\Delta({\bf k}, \omega_n)$, where  $\omega_n$ is
the Matsubara frequency. While $\Delta$ for conventional superconductors
is an even function of frequency,  the  class is one in which
$\Delta$ is odd in $\omega_n$ and, as a  consequence of the Pauli
principle, odd under parity as well:  $\Delta(- {\bf k}, \omega_n) =  -
\Delta({\bf k}, \omega_n)$. Thus, this new  class can have spin singlet
$p$-wave pairing in contrast to the triplet  $p$-wave pairing which
occurs in $^3$He with a gap which is even in  $\omega_n$.

In this paper we show that a stable odd (i.e. odd in $\omega_n$) singlet
pairing is unlikely to occur for a spin-independent effective potential,
e.g. a phonon interaction. This is because renormalization effects
reduce the dressed $p$-wave coupling below the threshold value for  $T_c
>0$, regardless how strong the bare coupling is. However, this
difficulty can be overcome if spin-dependent terms are added to the
interaction, such as may occur in high-$T_c$ superconductors  because of
antiferromagnetic fluctuations, or other strongly correlated systems.
Whether this situation is  realized in nature remains unclear.

We first consider a phonon model
\begin{equation}
H  = \sum_{{\bf k},\sigma}\epsilon_{{\bf k}} n_{{\bf k} \sigma} +  \sum_{{\bf
k},{\bf k}',
\sigma} g_{{\bf k}, {\bf k}'} \, c^{+}_{{\bf k}' \sigma} c_{{\bf k}  \sigma} (
a_{{\bf k} -
{\bf k}'} + a^{+}_{{\bf k}' - {\bf k}}) + \sum_{{\bf q}}  \omega_{{\bf q}}
N_{{\bf q}},
\end{equation}
where $  a_{{\bf k} } + a^{+}_{-{\bf k} }$ is the phonon
coordinate. At $T =  T_c$, the Eliashberg equations for spin singlet odd
$l$ pairing [e.g.  $\Delta_{{\bf k}}(\omega_n) = \Delta_1(\omega_n) \, {\hat
{\bf k} } \cdot  {\hat {\bf d}}\,$ for $p$-wave pairing] are:
\begin{equation}
\Delta_l(\omega_n) = \pi T \sum_{n'}{K_l(\omega_n -
\omega_{n'})\over{|Z(\omega_{n})| |\omega_{n'}|}}  \Delta_l(\omega_{n'})
\end{equation}
\begin{equation}
Z(\omega_n) = 1+{1\over{\omega_n}} \pi T
\sum_{n'}  K_0(\omega_n  -\omega_{n'}){\omega_{n'}\over{|\omega_{n'}|}}.
\end{equation}
The interaction kernels are defined by
\begin{equation}
K_l(\omega_n -\omega_{n'}) = N_0\int d\mu P_l(\mu) g^2_{\mu} {2
\Omega_{\mu}\over{(\omega_n - \omega_{n'})^2 + \Omega^2_{\mu}}}.
\end{equation}
We have set $|{\bf k}| = |{\bf k}'| = k_F$ and defined $\mu = {\bf k}
\cdot  {\bf k}'/k^2_F$. $P_l$ is the Legendre polynomial.

For $\Delta_l$ even in $\omega_n$, $T_c$ is nonzero regardless of  how
small $K_l$ is, so long as $K_l$ is positive (attractive), i.e. the
Cooper instability. As discussed in \cite{ba}, for the odd case, only
the  part of $K_l(\omega_n - \omega_{n'})$ which is odd in $\omega_n$
enters; this suppresses the density of states near the Fermi
surface by a factor $\omega^2_n$ (one power from $K$ and one from
${\Delta_ l}$) requiring a finite value of the interaction for $T_c$ to be
nonzero.

At first sight, it would appear that one could choose $g^2_0$ to be
small so that $Z \simeq 1$; then by increasing $g^2_{l>0}$ above
threshold, $T_c >0$ can be obtained. However, we note that the
derivability of $K$ from the phonon Hamiltonian requires, because of the
positivity of the phonon spectral function, that
\begin{equation}
K_{{\bf k} - {\bf k}'}(\omega_n - \omega_{n'}) = \sum_{l=0}^{\infty} (2l + 1)
P_l (\mu) K_l (\omega_n - \omega_{n'}) \geq 0.
\end{equation}
For example, if only $K_0$ and $K_1$ are non zero, then
\begin{equation}
|K_1| < |K_0|/3.
\end{equation}
In this case, $T_c \equiv 0$ since the effective
interaction $K^{eff}_1  = K_1/Z=K_1/(1+K_0)$ must be larger than $1$ for
$T_c > 0$ \cite{ba}. However, from Eq.\ (6) it follows that $K^{eff}_1 <
(K_0/3)/(1+K_0) < 1/3$ \cite {pba}.

To avoid this difficulty, we consider a  general electron-electron coupling. In
that case, the
low-energy behavior will be determined by an effective interaction which is
retarded and spin
dependent. Explicitly, we introduce a general spin- and frequency-dependent
coupling
\begin{equation}
\gamma(\alpha k; \beta k'|
\gamma p; \delta p') = \gamma^c(k -  p)\delta_{\alpha \beta}
\delta_{\gamma \delta} + \gamma^s(k - p)   \sigma^i_{\alpha \beta}
\sigma^i_{\gamma \delta},
\end{equation}
where $\alpha, \beta$ etc. are
spin indexes; $k,p$ etc. are 4-vectors, three of which are independent.

In this case, the Eliashberg equations in the spin singlet  $l$-wave
channel become
\begin{equation}
{\Delta_ l}(\omega_n) = - \pi T \sum_{n'} [
\gamma^c_l(\omega_n -  \omega_{n'}) - 3 \gamma^s_l(\omega_n -
\omega_{n'})]  {{\Delta_ l}(\omega_{n'})\over{|Z(\omega_{n})| |\omega_{n'}|}},
\end{equation}
\begin{equation}
Z(\omega_n)= 1+ \pi T \sum_{n'} [
\gamma^c_0(\omega_n -  \omega_{n'}) + 3 \gamma^s_0(\omega_n -
\omega_{n'})]  {\omega_{n'}\over{\omega_n |\omega_{n'}|}}.
\end{equation}
The change of sign of the $\gamma^s$ interaction in Eqs.\ (8,9)
provides the possibility of density and spin couplings adding in  the
pairing channel yet opposing each other in the normal self  energy
channel, so that, as we shall see below,  $Z$ remains $\sim 1$.

In addition, there is no apparent restriction on the $\gamma^{c, s}$
analogous to Eq.\ (5) for a general four-point vertex. Therefore, $T_c
> 0$ for odd ${\Delta_ l}$ may be realized for systems having strong
spin-dependent interactions.

For example, within the RPA for the Hubbard model,
\begin{equation}
\gamma^{c(s)}({\bf k} - {\bf k}', \omega_n - \omega_{n'}) = \pm   {U/2\over{ 1
\pm  U \chi_0({\bf k} - {\bf k}', \omega_n - \omega_{n'})}},
\end{equation}
where plus and minus signs correspond to charge ($c$) and spin ($s$)
channels. In the high-$T_c$ materials, it is observed experimentally
that $\gamma^s$ is enhanced for ${\bf k} - {\bf k}' \simeq Q = (\pm \pi,  \pm
\pi)$. This condition corresponds to backscattering. The $p$-wave
part of $\gamma^{s(c)}$ in 2D is
\begin{equation}
\gamma^{s(c)}_{1}(\omega_n - \omega_{n'}) = {1\over{\pi}}  \int_0^{\pi}
\cos{\theta}d\theta \, \gamma^{s (c)}({\bf k} - {\bf k}',  \omega_n -
\omega_{n'}),
\end{equation}
while the $s$-wave part is
\begin{equation}
\gamma^{s (c)}_{0}(\omega_n -\omega_{n'}) = {1\over{\pi}}
\int_0^{\pi}   d\theta \,    \gamma^{s(c)}({\bf k} - {\bf k}', \omega_n -
\omega_{n'}).
\end{equation}
Since $\gamma^s$ is negative for all
$\theta$ but is peaked for  $\theta \sim \pi$, where $ \cos {\theta} = -
1$, it follows that  $\gamma^s_1 > 0$ while $\gamma^s_0 < 0$. On the
other hand  $\gamma^c_0 > 0$ and we expect that  $\gamma^c_1$ is
small.

Here, because of the frequency summations in Eqs.(8-9), only the  odd in
$\omega_n, \omega_{n'}$ parts of  $\gamma^{c(s)}({\bf k} - {\bf k}',
\omega_n - \omega_{n'})$ enter the  Eliashberg equations. Now consider the
strong-correlation regime in the presence of the shadow upper and lower Hubbard
bands.
Then $\partial  \Sigma(\omega_n)/\partial  \omega_n > 0 $ over most of the
frequency
range up to $\omega_n \sim U$ \cite{KS}, in contrast to the conventional Fermi
liquid
in which  $\partial \Sigma(\omega_n)/\partial \omega_n < 0$. Consequently for
those
frequencies high enough to be relevant for odd pairing, $Z(\omega_n) = 1 -
\Sigma(\omega_n)/ \omega_n \sim 1 $.
This is sufficient  to have $Z$ of order unity while  still
having an attraction in the $p$-wave spin-singlet pairing  channel.

Under the circumstances being considered here, standard BCS $s$-wave singlet
pairing is
impossible because the interaction is repulsive in that channel.

We note that the condition $Z>1$ is required for stability. This has been
recently
reemphasized \cite{dol}. Contrary to the claim of this reference, however, our
discussion
obviously nowhere implies that $Z<1$ is a necessary condition for odd-frequency
pairing. In
fact, the arguments of Ref. 4 are irrelevant to the issue of odd-frequency
pairing.

We mention that a recent Monte Carlo simulation study for the
Cooper-pair $t$-matrix in the two-dimensional Hubbard model
\cite{bsw} contains results which indicate the possibility of odd-frequency
pairing.
These calculations show that the dominant singlet pair eigenvalues occur in the
$d_{x^2-y^2}$ (even-gap) and $p$-wave (odd-gap) channels.

In summary, we have shown that pairing by phonons is unlikely to  give
the odd gap singlet superconductor discussed in \cite{ba}. We  have
shown how a general electron-electron interaction can mediate such pairing and
have given a
concrete example of how this can work in the context of high-$T_c$
superconductivity, i.e. for
the Hubbard  model in 2D.

This work was partially supported by grant NSF-DMR 89-18307 (JRS), by
a J.R.~Oppenheimer Fellowship and the Department of Energy (AVB), by grant
NSF-DMR 89-06958 (EA), by the Advanced Studies Program of the Center
for Materials Science at Los Alamos National Laboratory and by grant NSF-DMR
91-18414 (PBA).

We thank D.J. Scalapino and P. W\"{o}lfle for valuable suggestions.

\end{document}